\documentclass[aps,prd,twocolumn,showpacs,preprintnumbers,amsmath,amssymb,floatfix,nofootinbib]{revtex4}
\usepackage{graphicx}

\usepackage{color}
\usepackage{colordvi}

\newcommand{\beqn}{\begin{eqnarray}}
\newcommand{\eeqn}{\end{eqnarray}}
\newcommand{\be}{\begin{equation}}
\newcommand{\ee}{\end{equation}}

\newcommand{\ba}{\begin{array}{c}}
\newcommand{\bat}{\begin{array}{cc}}
\newcommand{\ea}{\end{array}}

\newcommand{\bi}{\begin{itemize}}
\newcommand{\ei}{\end{itemize}}

\newcommand{\ket}{\,\rangle}
\newcommand{\bra}{\langle \,}

\newcommand{\Frac}[2]{\frac{\displaystyle #1}{\displaystyle #2}}

\newcommand{\cO}{{\cal O}}

\newcommand{\mF}{\mathcal{F}}

\newcommand{\mL}{\mathcal{L}}
\newcommand{\mM}{\mathcal{M}}

\newcommand{\lsim}{\stackrel{<}{_\sim}}

\begin{document}

%%\preprint{IFIC 09/??}
%%\preprint{SFB-CPP 09/??}
%%\preprint{PITHA 09/??}

%%%%%%%%%%%%%%%%%%%%%%%%%%%% TITLE %%%%%%%%%%%%%%%%%%%%%%%%%%%%%%%%%%%%%%%%%%

\title{Renormalization group  equations in resonance chiral theory }

\author{J.J. Sanz-Cillero}

%%%\email{cillero@ifae.es}

\affiliation{Grup de Fisica Teorica and IFAE,
Universitat Autonoma de Barcelona,
E-08193 Bellaterra (Barcelona), Spain
}

%%%\today

\begin{abstract}
The use of the equations of motion and meson field redefinitions
allows the development of a simplified resonance chiral theory lagrangian:
terms including resonance fields and a large number of derivatives
can be reduced into corresponding $\cO(p^2)$ resonance operators, containing the
lowest possible  number of derivatives.
This is shown by means of the explicit computation  of the pion vector form-factor
up to next-to-leading order in $1/N_C$.
The study of the renormalization group equations
for the corresponding couplings demonstrates the existence of an infrared
fixed point in the resonance theory.
The possibility of developing a perturbative $1/N_C$ expansion
in the slow running region around the fixed point is shown here.
\end{abstract}

%\vskip .5cm


\maketitle

%%%%%%%%%%%%%%%%%%%%%%%%%%%%%%%%%%%%%%%%%%%%%%%%%%%%%%%%%%%%%%%%%%%%%%%%%%%%%%

\section*{$1/N_C$ expansion in resonance chiral theory}

Resonance chiral theory (R$\chi$T) is a description  of the Goldstone-resonance
interactions within a chiral invariant framework~\cite{RChTa,RChTb}.
The pseudo-Goldstone fields $\phi$ are introduce through the exponential  realization
${    u(\phi) =\exp{\left(i \phi/\sqrt{2} F\right)}      }$.  The standard effective field theory
momentum expansion is not valid in the presence of heavy resonance states
and an alternative perturbative counting is required.
R$\chi$T  takes then the formal $1/N_C$ expansion as a guiding principle~\cite{NC}:
at leading order (LO) the interaction terms
in the lagrangian with a number $k$ of meson fields (and their corresponding couplings)
scale as $\sim N_C^{1-\frac{k}{2}}$~\cite{NC}. For instance,
the resonance masses are counted as $\cO(N_C^0)$, the three-meson vertex operators
are $\cO(N_C^{-1/2})$, etc.
The subdominant terms in the lagrangian will have then subleading $1/N_C$ scalings
with respect to these ones.
If our action is now arranged according to the number
of resonance fields in the operators, one has
\begin{eqnarray}
 \mathcal{L}_{\mathrm{R}\chi\mathrm{T}}&=& \mathcal{L}^{\mathrm{GB}} + \mathcal{L}^{R_i} + \mathcal{L}^{R_iR_j} + \mathcal{L}^{R_iR_jR_k}
 +\dots \,,
 \label{lagr}
\end{eqnarray}
where the resonance fields $R_i$ are classified in $U(n_f)$ multiplets,
with $n_f$ the number of light quark flavours.

A priori,  ${\cal L}_{\mathrm{R}\chi\mathrm{T}}$ might contain
chiral tensors of arbitrary order. However, for most phenomenological
applications, terms with a large number of derivatives tend  to violate the asymptotic
short-distance behavior of QCD Green Functions and form factors~\cite{PI:08}.
Likewise, it is possible to prove that in the chiral limit
the most general $S-\pi\pi$ interaction is provided  by the
operator of lowest order in derivatives~\cite{cd+EoM}.
A similar proof can be derived for the  $V-\pi\pi$ vertex~\cite{prepara}

The operators of the leading R$\chi$T lagrangian
without resonance fields are those from $\chi$PT at $\cO(p^2)$~\cite{chpt},
\begin{eqnarray}
 \mathcal{L}_{\rm LO}^{\mathrm{GB}} \,\,
 =\,\, \Frac{F^2}{4}\, \langle u_\mu u^\mu +\chi_+\rangle
 \,.
\end{eqnarray}
The Goldstones fields, given by $u(\phi)$, enter in the lagrangian through
the covariant tensors
${      u_\mu=i\{
u^\dagger (\partial_\mu - i r_\mu) u- u (\partial_\mu - i\ell_\mu) u^\dagger\}       }$
and $       \chi_\pm=u^\dagger \chi u^\dagger \pm u \chi^\dagger u      $,
with $\ell_\mu$, $r_\mu$ and $\chi$ respectively the left-current, right-current
and scalar-pseudoscalar density sources~\cite{RChTa,op6-chpt}. Likewise, it is
convenient to define
$f_\pm^{\mu\nu}=u F_L^{\mu\nu} u^\dagger \pm u^\dagger F_R^{\mu\nu} u$,
with $F_{L,R}^{\mu\nu}$ the left and right field strength tensors~\cite{RChTa,op6-chpt}.

In the case of the vector multiplet, one has at LO in $1/N_C$
the operators~\cite{RChTa}
\begin{align}
\mathcal{L}_{\rm LO}^V &=\, \frac{F_V}{2\sqrt{2}} \bra V_{\mu\nu} f^{\mu\nu}_+ \ket \,
+\, \frac{i\, G_V}{2\sqrt{2}} \bra V_{\mu\nu} [u^\mu, u^\nu] \ket
\, ,  \label{1Rlagrangian}
\end{align}
where  the antisymmetric tensor field $V^{\mu\nu}$ is used in R$\chi$T
to describe the spin--1 mesons~\cite{RChTa,RChTb,chpt}, with the
kinetic and mass terms,
\begin{align}
\mathcal{L}^V_{\rm Kin} &=\, -\, \Frac{1}{2}
\bra V_{\lambda\nu} \nabla^\lambda\nabla_\rho V^{\rho \nu}\ket
\,+\,\Frac{1}{4}M_V^2\bra V_{\mu\nu} V^{\mu\nu}\ket \, .
\label{eq.Kin}
\end{align}
The covariant derivative is defined through  $\nabla_\mu X =\partial_\mu X
+[\Gamma_\mu , X]$,  with the chiral connection
${   \Gamma_\mu=\frac{1}{2}\{ u^\dagger (\partial_\mu - i r_\mu) u
+ u (\partial_\mu - i\ell_\mu) u^\dagger\}     }$.
Other works have widely studied alternative representations of the
vector mesons such as general four-vector formalisms~\cite{spin1,Pallante},
the gauged chiral model~\cite{Schechter85,Donoghue89}
or  the hidden local symmetry framework~\cite{spin1,HLS,Wilson}.

The naive dimensional analysis of the operators tells us that
the tree-level LO amplitudes will scale like $\mM\sim p^2$ in the external
momenta $p$. At one loop,  higher power corrections
$\mM\sim p^4 \ln(-p^2)$ are expected to arise.
These logs will come together with ultraviolet (UV)  divergences $\lambda_\infty p^4$,
requiring  new operators   subleading  in $1/N_C$,
with a larger number of derivatives with respect to the leading order ones.
These $\cO(p^4)$ corrections look, in principle, potentially dangerous if the
momenta become of the order of the resonance masses.
Since there is no characteristic scale $\Lambda_{\rm R\chi T}$ that suppresses them
for $p\ll \Lambda_{\rm R\chi T}$,
they could become as important as the $\cO(p^2)$ leading order contributions.

In the present case of the $\pi\pi$
vector form-factor (VFF),  in order to fulfill the one-loop renormalization one needs
the subleading operators~\cite{L9}
\begin{align}
\mathcal{L}^{\rm GB}_{\rm NLO} &=\,
-\, i\, \widetilde{L}_9\bra f_+^{\mu\nu} u_\mu u_\nu\ket \, ,
\nonumber\\
\mathcal{L}^V_{\rm NLO} &=\,
X_Z\bra V_{\lambda\nu}\nabla^\lambda\nabla_\rho \nabla^2 V^{\rho\nu}\ket
\,+\,  X_F\bra V_{\mu\nu}\nabla^2 f_+^{\mu\nu}\ket
\nonumber\\
& \qquad + \,2\, i\, X_G    \bra V_{\mu\nu} \nabla^2 [u^\mu, u^\nu] \ket
\,.
\label{eq.lagrNLO}
\end{align}
However,  the $\mathcal{L}^V_{\rm NLO}$
couplings $X_{Z,F,G}$ are not physical by themselves:
it is impossible to fix them univocally from the experiment.
Indeed, since these subleading $\mL^V_{\rm NLO}$ operators are proportional to the equations
of motion, one finds that
$\mathcal{L}^V_{\rm NLO}$ can  be fully transformed
into the $M_V$, $F_V$, $G_V$ and $\widetilde{L}_9$ terms  and
into other operators that  do not contribute
to the VFF by means of meson field redefinitions~\cite{L9,Tesis}.
Furthermore, higher derivative resonance operators that could contribute to the
VFF at tree-level can be also removed from the lagrangian in the same way~\cite{prepara}.

One of the aims of this article is to show how the potentially dangerous higher
power corrections arising at next-to-leading order (NLO)~\cite{NLO,L10}
actually correspond to a slow logarithmic
running of the couplings of the LO lagrangian.
We will make use of the equations
of motion of the theory and meson field redefinitions  to remove
analytical corrections going like higher powers
of the momenta. This  leaves just the problematic log  terms $p^4\ln(-p^2)$,
which will be  minimized by means of the renormalization group equations
and transformed into a logarithmic running of $M_V$, $F_V$, $G_V$ and $\widetilde{L}_9$.

%\vspace*{-1cm}
\section*{The pion vector form-factor}

In order to exemplify the procedure, the rest of the article is devoted to
a thorough study of the pion vector form-factor  in the chiral limit:
\be
\label{eq.def-VFF}
\bra \pi^-(p_1) \pi^0(p_2)|\bar{d}\gamma^\mu u|0\ket \,\,=\,\,
\sqrt{2}\,(p_1^\mu -p_2^\mu)\,\, \mF(q^2)\, ,
\ee
with $q\equiv p_1+p_2$.

\begin{figure}
\begin{center}
\includegraphics[angle=0,clip,width=7cm]{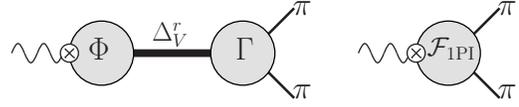}
\caption{{\small 1PI-topologies contributing to the  pion VFF.
 }}
 \label{fig.1PI}
\end{center}
\end{figure}

The renormalized amplitude shows the following general structure in terms of
renormalized vertex functions and the renormalized vector correlator,
\be
\label{eq.NLO-VFF}
\mF(q^2) = \mF(q^2)_{\rm 1 PI} + \Frac{\Phi(q^2) \Gamma(q^2)}{F^2}
\Frac{q^2}{M_V^2 -q^2-\Sigma(q^2) }\ ,
\ee
with $\Sigma(q^2)$ the vector self-energy, and $\mF(q^2)_{\rm 1 PI}$, $\Phi(q^2)$
and $\Gamma(q^2)$ being provided, respectively,
by the 1--particle-irreducible (1PI) vertex functions
for $J_V^\mu\to\pi\pi$, $J_V^\mu\to V$ and $V\to\pi\pi$  (Fig.~\ref{fig.1PI}).
Thus, at large $N_C$, R$\chi$T yields for the VFF,
\be
\label{eq.LO-VFF}
\mF(q^2) = 1 + \Frac{F_V G_V}{F^2}
\Frac{q^2}{M_V^2 -q^2 }\ .
\ee

Although QCD contains an infinite number of hadronic states,
only a finite number of them is considered for most phenomenological
analyses~\cite{PI:08}.
We will include in the R$\chi$T just the lightest mesons (Goldstones and vectors).
Likewise, only the lowest threshold contributions are taken into account in this work
--the massless two-Goldstone cut-- and   loops from higher cuts will be assumed to
be renormalized in a $\mu$--independent scheme, such that they decouple
as far as the total energy remains below their production threshold
(see for instance Appendix C.2 in Ref.~\cite{L10}).
In general, all the considerations along the paper will be restricted
to this range. Only at the end we will allow a small digression
about speculations and results for our form-factor calculation
in the high-energy limit.

The one-loop calculation produces a series of ultraviolet divergences
that require of subleading operators in $1/N_C$ ($X_Z$, $X_F$, $X_G$, $\widetilde{L}_9$)
to fulfill the renormalization of the vertex functions~\cite{L9,Tesis}:
\begin{eqnarray}
\Sigma(q^2)&=& -2 q^4 X_Z
- \Frac{n_f}{2}\Frac{2 G_V^2 }{F^2}\Frac{q^4}{96 \pi^2 F^2}
\ln\Frac{-q^2}{\mu^2}\,,
\nonumber\\
\Gamma(q^2)&=&\, - 4 \sqrt{2} X_G q^2
\nonumber\\
&&\hspace*{-1.7cm}  + G_V\left[  1
- \Frac{n_f}{2} \left(1-\Frac{G_V^2}{F^2}\right)
\Frac{q^2}{96 \pi^2 F^2} \ln\Frac{-q^2}{\mu^2}  +\Delta_t(q^2)\right]  ,
\nonumber\\
\Phi(q^2)&=& F_V  - 2\sqrt{2} X_F  q^2
-  \Frac{n_f}{2} \Frac{2 G_V}{F^2}  \Frac{q^2}{96\pi^2}
\ln\Frac{-q^2}{\mu^2}  \, ,
\nonumber\\
\mF(q^2)_{\rm 1 PI}&=&\,1 +\Frac{ 2 q^2 \widetilde{L}_9 }{F^2} +\Delta_t(q^2)
\nonumber\\
&&\hspace*{0.5cm}
- \Frac{n_f}{2} \left(1-\Frac{G_V^2}{F^2}\right)
\Frac{q^2}{96 \pi^2 F^2} \ln\Frac{-q^2}{\mu^2}   \, ,
\label{eq.vertex1}
\end{eqnarray}
being $n_f$ the number of light flavours and $\Delta_t$
the finite and $\mu$--independent contribution from the triangle diagram
that contains the $t$--channel exchange of a vector meson,
\be
\Delta_t(q^2) = \Frac{n_f}{2}\Frac{2G_V^2}{F^2}\Frac{M_V^2}{16\pi^2 F^2}
\hat{\Delta}_t(q^2/M_V^2)\,\, ,
\ee
with
%%%\begin{eqnarray}
%%%\hat{\Delta}_t(x)\, &=&\, \left[\mbox{Li}_2(1+x)-\mbox{Li}_2(1)\right]\, \left(\Frac{1}{x^2}
%%%+\Frac{5}{2 x} +1\right)
%%%\nonumber\\
%%%&&\qquad \,+\, \ln(-x)\, \left(\Frac{1}{x}+2\right) -\Frac{1}{x} -
%%%\Frac{9}{4}\, ,
%%%\end{eqnarray}
$\hat{\Delta}_t(x)= \left[\mbox{Li}_2(1+x)-\mbox{Li}_2(1)\right]
\left(\frac{1}{x^2} +\frac{5}{2 x} +1\right)+\ln(-x)\,
\left(\frac{1}{x}+2\right) -\frac{1}{x} - \frac{9}{4}$,
vanishing at zero
like $\hat{\Delta}_t=-\frac{1}{12}x\ln(-x) + \frac{35}{72}x+\cO(x^2)$ and growing
for large $x$ like a double log, $\hat{\Delta}_t\sim -\frac{1}{2}\ln^2 |x|$.
%%%It has
%%%slow logarithmic growing  with $\Delta_t(1)\simeq 0.6$,
%%%$|\Delta_t|<1$ for $|x|<2$,$|\Delta_t|<3$ for $|x|<10$,$|\Delta_t|<9$ for $|x|<100$...
%%In general we will find
%%this kind of finite double logs in the triangle diagrams with the exchange of a heavy particle.
%%To minimize this double logs, maybe one should develop some kind of
%%Sudakov-like resummation.
%%This remains nevertheless out of the scope of this article.
For the energies we are going to study ($|q^2|\lsim 1$~GeV$^2$),
it will have little numerical impact.

The couplings that appear in the finite vertex  functions in~(\ref{eq.vertex1})
are the renormalized ones. The NLO running of $G_V(\mu)$
induces then a residual $\mu$--dependence in~(\ref{eq.vertex1})
at next-to-next-to-leading order (NNLO)
which  allows us to use the renormalization group techniques ro resum harmful
large radiative corrections.
However, the NLO operators $X_{Z,F,G}$ from~(\ref{eq.lagrNLO})
are found to be proportional to the equations of motion~\cite{L9,Tesis}.
The physical meaning of this is that these parameters can be never
extracted from the experiment in an independent way. The amplitudes rather depend
on effective combinations of them and other couplings.
Thus, it is possible to transform the renormalized part of these operators
into the $M_V$, $F_V$, $G_V$ and $\widetilde{L}_9$ operators and
other terms that do not contribute to the amplitude
by means of a convenient meson field redefinition
$V\longrightarrow  V\,+\,\xi(X_Z,X_F,X_G)$~\cite{L9,Tesis}:
\begin{eqnarray}
&&X_{Z,F,G}  \stackrel{\xi}{\longrightarrow}   0\, \,,
\nonumber\\
&&\widetilde{L}_9   \stackrel{\xi}{\longrightarrow}
\widetilde{L}_9 + \left( \sqrt{2}  X_F G_V
\hspace*{-0.075cm}+ \hspace*{-0.05cm}2\sqrt{2} F_V X_G
\hspace*{-0.05cm}- \hspace*{-0.05cm}X_Z F_V G_V
\right) ,
\nonumber\\
&&F_V   \stackrel{\xi}{\longrightarrow}
F_V + \,\left( 2 X_Z F_V M_V^2 - 2 \sqrt{2} X_F M_V^2 \right) \,,
\nonumber\\
&&G_V   \stackrel{\xi}{\longrightarrow}
G_V + \,\left( 2 X_Z G_V M_V^2 - 4 \sqrt{2} X_G M_V^2 \right) \,,
\nonumber\\
&&M_V^2   \stackrel{\xi}{\longrightarrow}
M_V^2 + 2 \, \,X_Z M_V^4 \,.
\end{eqnarray}
Hence, it is possible then to consider a suitable shift  that removes the
renormalized operators $X_{Z,F,G}$ from the lagrangian, encoding their information and running in the remaining
$\widetilde{L}_9$, $F_V$, $G_V$ and $M_V$.
Although this transformation $\xi$ depends on the renormalization scale $\mu$ (as it depends on the renormalized $X_{Z,F,G}$), the resulting theory
is still   equivalent to the original one.
The redundant parameters $X_{Z,F,G}$  are removed for every $\mu$ from
the vector self-energy and  vertex functions in~(\ref{eq.vertex1}),
inducing in the remaining couplings a running ruled by the renormalization
group equations~(RGE),
\begin{eqnarray}
\label{eq.running}
\Frac{1}{M_V^2}\Frac{\partial M_V^2}{\partial\ln\mu^2} &=&  \Frac{n_f}{2}
\Frac{2 G_V^2}{F^2}\Frac{M_V^2}{96\pi^2 F^2}  \, ,
\\
\Frac{\partial G_V}{\partial \ln\mu^2}
&=&  G_V\,\, \Frac{n_f}{2}\Frac{M_V^2}{96\pi^2 F^2} \left( \Frac{ 3 G_V^2}{F^2}-1\right)\, ,
\nonumber\\
\Frac{\partial F_V}{\partial\ln\mu^2} &=& 2 \,G_V\,
\Frac{n_f}{2}\Frac{M_V^2}{96\pi^2 F^2} \left( \Frac{F_V G_V}{F^2}-1\right) \, ,
\nonumber\\
\Frac{\partial \widetilde{L}_9}{\partial\ln\mu^2}
&=&
\Frac{n_f}{2}\Frac{1}{192 \pi^2}\,
\left( \Frac{F_V G_V}{F^2} -1\right)\,
\left(1- \Frac{3 G_V^2}{F^2}\right)\, .
\nonumber
\end{eqnarray}

If one now takes the VFF expression given by~(\ref{eq.NLO-VFF})
and (\ref{eq.vertex1})  and sets $\mu^2=Q^2$ (with $Q^2\equiv -q^2$),
it gets the simple form,
\begin{eqnarray}
\label{eq.VFF}
&&\mF(q^2) = \,\, -\,\, \Frac{ 2 \,Q^2 \widetilde{L}_9(Q^2)}{F^2}
 \\
&&  \hspace*{-0.4cm}+
\left[1+\Delta_t(q^2)\right]\,\,
\left[1- \Frac{F_V(Q^2) G_V(Q^2)}{F^2}\Frac{Q^2}{M_V^2(Q^2)\,  +\,  Q^2}
\right] \, \, ,
\nonumber
\end{eqnarray}
with the evolution of the couplings with $Q^2$ prescribed by the
RGE~(\ref{eq.running}).  Notice that if the subleading
terms $\widetilde{L}_9$ and $\Delta_t(q^2)$ are dropped, one is left with
the resummed expression at leading log for the LO form-factor~(\ref{eq.LO-VFF}).
The residual NNLO dependence could be estimated by
varying $\mu^2$ around $Q^2$, in the range between $Q^2/2$ and $2 Q^2$, as
it is often done in RGE analysis.
In this scheme,  $M_V$ would be related to the pole mass through
${     M_{V,\, pole}^2
=M_V^2(\mu)+\frac{n_f}{2}\frac{2 G_V^2}{F^2}\frac{M_V^4}{96\pi^2 F^2}
\ln{\frac{M_V^2}{\mu^2}}=M_V^2(M_V)      }$.

%%%%These and the  low energy expansion of $\Delta_t(q^2)$
%%%%ensures the recovery of the
%%%%$\chi$PT running when $Q^2\to 0$.
%%%%
%%%%For instance, at $\cO(p^4)$ one recovers the log dependence,
%%%%\be
%%%%\lim_{Q^2\to 0}\,\,  \Frac{d}{d\ln Q^2}\left[\Frac{\mF(Q^2)-\mF(1)}{Q^2}\right]\,
%%%%\, =\,\,\Frac{\Gamma_9}{16\pi^2 F^2}\, ,
%%%%\ee
%%%%with $\Gamma_9= n_f/12$ for $\chi$PT with $n_f$ light quark flavours.
%%%%

\begin{figure}[!t]
\begin{center}
\includegraphics[angle=0,clip,width=8.5cm]{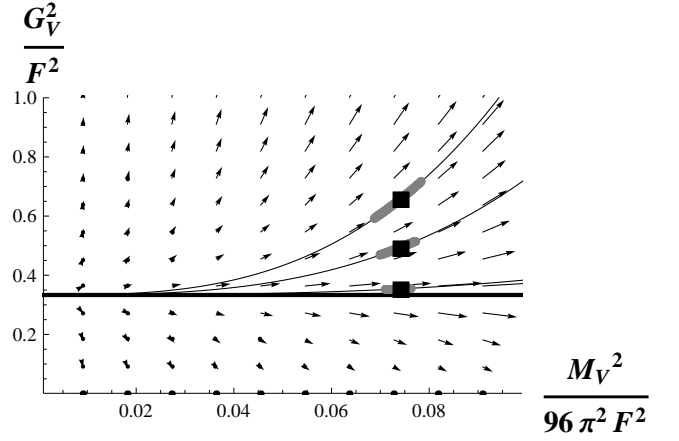}
\caption{{\small
Renormalization group flow for $M_V^2$ and $G_V^2$. The points $M_V(\mu_0)=775$~MeV and
$G_V(\mu_0)=75,\, 65,\, 55$~MeV are plotted with filled squares, together with their
trajectories for ${  n_f=2  }$ (thin black lines).
For illustrative purposes,
and assuming those as
initial conditions for $\mu_0=770$~MeV, we also show their running between $\mu=500$~MeV
and $\mu=1$~GeV (thick gray lines).  The
horizontal line represents the $G_V$--fixed point at $G_V^2=F^2/3$.
 }}
\label{fig.MVGV}
\end{center}
\end{figure}

The first two RGE refer to $M_V$ and $G_V$ and form a closed system with the
trajectories given by
\be
G_V^2\,\,=\,\, \Frac{F^2}{3}\,\left(\,1\, + \, \kappa^3\, M_V^6 \right)\, ,
\ee
with $\kappa$ an integration constant. It leads to the solution
\be
\Frac{1}{M_V^2} \, +\, \kappa \, f(\kappa M_V^2)\, \, =\, \,
-\,\Frac{2}{3}\, \Frac{n_f}{2}\Frac{1}{96\pi^2 F^2}\,
\ln\Frac{\mu^2}{\Lambda^2}\, ,
\label{eq.solMV}
\ee
with $f(x)= \frac{1}{6}\ln\left(\frac{x^2+ 2 x +1}{x^2 -x +1}\right)
- \frac{1}{\sqrt{3}}\arctan\left(\frac{ 2x-1}{\sqrt{3}}\right)-\frac{\pi}{6\sqrt{3}}
=\cO(x)$, and $\Lambda$ an  integration constant. Since
$-\frac{2\pi}{3\sqrt{3}}\leq f(x)\leq 0$, the term $\kappa f(\kappa M_V^2)$
in~(\ref{eq.solMV}) becomes negligible for very low momentum,
$\mu\ll \Lambda$, producing a logarithmic running.
The parameters $M_V$ and $G_V$ show then an infrared fixed point at
$M_V=0$ and $G_V=F/\sqrt{3}$.
The corresponding flow diagram  is shown in Fig.~\ref{fig.MVGV}.
The same happens for $F_V$ and $\widetilde{L}_9$,
which freeze out when $\mu\to 0$. $F_V$ tends to the infrared fixed point $\sqrt{3} F$
(and hence $F_V G_V\stackrel{\mu\to 0}{\longrightarrow} F^2$)
and $\widetilde{L}_9(\mu)$ goes to a constant value $\widetilde{L}_9(0)$.
%
%%%Thus, if $F_V=3 G_V=\sqrt{3} F$ for some $\mu_0\neq 0$ then $\widetilde{L}_9$
%%%and the two
%%%resonance couplings  result $\mu$--independent and the  $M_V$ running is given by
%%%Eq.~(\ref{eq.solMV}) with $\kappa=0$.

An analogous renormalization group analysis of the fixed points was also performed
in Ref.~\cite{Wilson} within a Wilsonian approach in the Hidden Local Symmetry
framework~\cite{HLS}.

\section*{A digression on high-energy constraints}

Although the present computation is only strictly valid below the first
two-meson   threshold with at least one resonance (since these channels
were not included here), one is allowed to speculate about the high-energy
behaviour of our expression~(\ref{eq.VFF}).

It is remarkable that the value of the resonance
couplings at the infrared fixed  point, $F_V G_V=F^2$ and $3 G_V^2=F^2$,
coincides with those obtained if one demands at large--$N_C$ the proper
high energy behaviour of, respectively,
the VFF~\cite{L9,PI:08} and the partial-wave scattering amplitude~\cite{scat}.

Likewise, it is also interesting to note that
the requirement that our one-loop form factor~(\ref{eq.VFF})
vanishes when $Q^2\to\infty$~\cite{PI:08,BrodskyLepage,NLOsatura}
leads to these same solutions:   the constraints
$F_V G_V=F^2$ and $3 G_V^2 =F^2$ are required to freeze out the running of
$\widetilde{L}_9$ and $F_V G_V$ and to kill the $q^2 \ln(-q^2)$ and $q^0 \ln(-q^2)$ short-distance
behaviour;  additionally, $\widetilde{L}_9=0$ is needed  in order to remove
the remaining $\cO(q^2)$ terms at $q^2\to\infty$.

The reason for this interplay between short distances and fixed points
is that in our case the massless logarithms come always together with the
UV--divergence $\lambda_\infty$ in the form
$\left[\lambda_\infty +\ln{\frac{Q^2}{\mu^2}}\right]$.  In similar terms,
these logs are related to the one-loop spectral function Im$\mF(q^2)$.
When only the two-Goldstone channel  $\phi\phi$  is open,
in the chiral limit,  the optical theorem states
\begin{eqnarray}
\mbox{Im}\mF_{_{\pi\pi}} &=& \sum_{\phi\phi} T^*_{_{\pi\pi\to\phi\phi}}\,\,
\mF_{_{\phi\phi}}
\nonumber \\
&\stackrel{q^2\to\infty}{=}&
\Frac{n_f}{2}\, \, \left[\Frac{q^2}{96 \pi F^2}\left(1- \Frac{3 G_V^2}{F^2}\right)
+\cO(q^0)\right]
\nonumber\\
&&\times \,\,
\left[\left(1-\Frac{F_V G_V}{F^2}\right) +\cO(q^{-2})\right]\, ,
\end{eqnarray}
with $\mF_{_{\phi\phi}}$ the vector form-factor with $\phi\phi$ in the final state
and $T_{_{\pi\pi\to\phi\phi}}$ the $I=J=1$ partial-wave scattering amplitude.
If the VFF spectral function is demanded to vanish at high energies then one necessarily
needs $1-\frac{3 G_V^2}{F^2}=0$ and $1-\frac{F_V G_V}{F^2}=0$.  These conditions
eliminate  the  $\cO(q^2 \ln(-q^2))$ and $\cO(q^0\ln(-q^2))$
logarithms and their accompanying $\cO(q^2)$ and $\cO(q^0)$ UV--divergences.
Regarding the tree-level contribution to the VFF,
$\mF(q^2)^{^{\rm tree}}\stackrel{q^2\to\infty}{=} \frac{2 \widetilde{L}_9}{F^2} q^2
+ \left(1- \frac{F_V G_V}{F^2}\right)+\cO(q^{-2})$,
one finds then that there is no running for $\widetilde{L}_9$ nor for $F_V G_V$.
The freezing in the running of the remaining combination, $G_V^2$, is due to
the $\cO(q^0)$ behaviour of  the one-loop $T_{_{\pi\pi\to\pi\pi}}$
spectral function  at $q^2\to\infty$,
\begin{eqnarray}
\mbox{Im}T_{_{\pi\pi\to\pi\pi}}&=&
\sum_{\phi\phi} |T_{_{\pi\pi\to\phi\phi}}|^2
\\
&\stackrel{q^2\to\infty}{=}&
\Frac{n_f}{2}\, \, \left[\Frac{q^2}{96 \pi F^2}\left(1- \Frac{3 G_V^2}{F^2}\right)
+\cO(q^0)\right]^2 \, ,
\nonumber
\end{eqnarray}
after imposing the former constraint $1-\frac{3 G_V^2}{F^2}=0$.
The $\cO(q^2)$ logs and the accompanying  UV--divergences
are absent in the $\pi\pi$ partial-wave amplitude. Hence, no running is induced
in the corresponding $\cO(q^2)$ combination of couplings that
is  relevant for the scattering  amplitude, which in R$\chi$T happens to be $G_V^2$.

\begin{figure}[!t]
\begin{center}
\includegraphics[angle=0,clip,width=8.5cm]{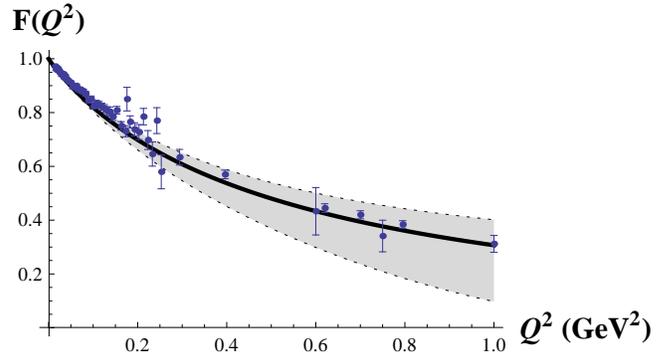}
\caption{{\small
Illustrative comparison of the VFF at NLO and euclidean data
($Q^2=-q^2>0$)~\cite{Amendolia}.
We have used ${  M_V(\mu_0)=775   }$~MeV,  $F_V(\mu_0) G_V(\mu_0) = 3\, G_V(\mu_0)^2 =F^2$
and $\widetilde{L}_9(\mu_0)=0$ for $\mu_0=770$~MeV (solid line).
The mild relevance of $G_V$ within the loops
in the euclidean range is represented by the gray band,
which shows the VFF for a large variation of the input $G_V(\mu_0)^2$,
in the range from zero up to $F^2$, while   $F_V(\mu_0) G_V(\mu_0)$, $M_V(\mu_0)$
and $\widetilde{L}_9(\mu_0)$ are kept the same as before.
%%
%%We have used $M_V=776$~MeV, $G_V=69$~MeV, $F_V=113$~MeV and $\widetilde{L}_9=0$
%%for $\mu=770$~MeV (solid line).
}}
\label{fig.euclid}
\end{center}
\end{figure}

In Fig.~\ref{fig.euclid}, the   VFF~(\ref{eq.VFF})
is compared  with   euclidean data in the range
$Q^2=0\,\,$--$\,\, 1$~GeV$^2$~\cite{Amendolia},
with the values $M_V=775$~MeV, $F_V = 3 G_V=\sqrt{3} F$, $\widetilde{L}_9=0$
for $\mu_0=770$~MeV.
Although our expression neglects contributions  from
higher channels, these values produce  a fair agreement with the data
in Fig.~\ref{fig.euclid}.
Nevertheless,
the non-zero pion mass is responsible of a 20\%
decreasing in the $\rho$ width~\cite{rho-width}  and
an accurate description of both the spacelike and timelike data requires
the consideration of the pseudo-Goldstone masses.
The residual NNLO  dependence was estimated by varying the scale $\mu^2$ between
$Q^2/2$ and $2 Q^2$ in~(\ref{eq.vertex1}), finding a shift of less than 0.3\%
for the inputs under consideration.

%In any case, the interplay between the infrared RGE
%flow and short-distance constraints
%remains unclear and it should require further studies in future works.

\section*{Perturbative regime in the $1/N_C$ expansion}

Independently of any possible high energy matching~\cite{PI:08,NLOsatura},
what becomes clear from the RGE analysis is the existence
of a region in the RGE space of parameters (around the infrared fixed point at
$\mu\to 0$) where the loops produce small logarithmic corrections.
Although we start with a formally well defined $1/N_C$ expansion,
this is the range where the perturbative description actually makes sense
for the renormalized  R$\chi$T amplitude. In an analogous way, although
the fixed order perturbative QCD cross-section calculations
are formally correct for arbitrary $\mu$  (and independent of it),
perturbation theory can only be applied at high energies.
In our case, the parameter that actually rules
the strength of the resonance-Goldstone interaction
in the RGE of Eq.~(\ref{eq.running}) is
\be
\alpha_V\,\, = \,\,  \Frac{n_f}{2}
\Frac{2 G_V^2}{F^2}\Frac{M_V^2}{96\pi  F^2} \, ,
\label{eq.width}
\ee
which goes to zero as $\mu\to 0$.
Thus, although the formal expansion parameter of the   theory is $1/N_C$,
this is the actual quantity that appears in the calculation suppressing the subleading
contributions.   Since at lowest order   $\alpha_V$ is just the
ratio of the vector width and mass, $\Gamma_V/  M_V\simeq 0.2$~\cite{rho-width,anchura-Jorge},
a $1/N_C$ expansion of R$\chi$T is meaningful
as far as the concerning resonance is narrow enough (as it happens here).

In the case of broad states or more complicate processes,
the identification of the parameter that characterizes
the strength of the interaction can be less intuitive. Nonetheless, perturbation
theory will be meaningful in R$\chi$T
as far as there is an energy range where this strength-parameter becomes small,
bringing along a slow running for the resonance couplings in the problem.

\section*{Conclusions}

Although, {\it a priori}, R$\chi$T needs of higher derivative operators
at NLO, not all the new couplings are physical.
The combination of meson field redefinitions and renormalization group equations
allows us to develop an equivalent theory without redundant operators
and where undesirable higher power corrections
are absent.

The study of the running of the couplings entering in the pion vector form-factor
shows the existence of an infrared fixed point. The couplings enjoy
a slow logarithmic running in the low-energy region around $\mu\to 0$,
where the resonance-Goldstone strength parameter $\alpha_V$ is small enough.
It is in this range of momenta that perturbation theory in  $1/N_C$
makes sense for R$\chi$T.

The physical amplitudes are then understood  in terms of renormalized
resonance couplings which evolve   with $\mu$ in the way prescribed
by the RGE.
A perturbative description of the observable  will be possible
as far as the loops keep their running slow.

These considerations are expected to be relevant for the study of
other QCD matrix elements.  In particular, they may play
an important role in the case of scalar resonances. The
width and radiative corrections are usually  rather sizable in the spin--0 channels.
The possible presence of fixed points and slow--running regions in other
amplitudes (e.g. the pion scalar form-factor)   will be studied in future analyses.

%%\vspace*{0cm}

\acknowledgments

%\vspace*{-0.1cm}
This work has been supported in part by
CICYT-FEDER-FPA2008-01430, SGR2005-00916, the Spanish Consolider-Ingenio 2010
Program CPAN (CSD2007-00042), the Juan de la Cierva program and
the EU Contract No. MRTN-CT-2006-035482
(FLAVIAnet).
I would like to thank V.~Mateu, I.~Rosell, P.~Ruiz-Femen\'\i a and S.~Peris for their
help and comments on the manuscript.

%\vspace*{0.1cm}

%%%%%%%%%%%%%%%%%%%%%%%%%%% REFERENCES %%%%%%%%%%%%%%%%%%%%%%%%%%%%%%%%%%%

\end{document}